\documentclass[letter]{aa} 

\newcommand{\xmm}{{\em XMM-Newton}}

\newcommand{\rhoOph}{$\rho$~Ophiuchi}
\newcommand{\pn}{{\em pn}}
\usepackage{graphicx}
\usepackage{txfonts}
\begin{document} 
\title{Smooth X-ray variability from $\rho$ Ophiuchi A+B.}
\subtitle{A strongly magnetized primary B2 star?}

\author{Ignazio Pillitteri \inst{1}
\and
Scott J. Wolk
\inst{2}
\and
Alyssa Goodman\inst{2}
\and
Salvatore Sciortino\inst{1}}
\institute{INAF-Osservatorio Astronomico di Palermo\\
Piazza del Parlamento 1, 90134 Palermo, Italy
\email{pilli@astropa.inaf.it} 
\and
Smithsonian-Harvard Center for Astrophysics \\
60 Garden St, Cambridge MA, 02138 USA 
}

\date{Received; accepted }

\abstract{
X-rays from massive stars are ubiquitous yet not clearly understood.
In an \xmm\ observation devoted to observe the first site of 
star formation in the \rhoOph\ dark cloud, we detect smoothly variable X-ray emission 
from the B2IV+B2V system of \rhoOph. Tentatively we assign the emission
to the primary component. The light curve of the \pn\ camera
shows a first phase of low, almost steady rate, then a rise phase of duration 
of 10 ks, followed by a high rate phase. 
The variability is seen primarily in the band 1.0-8.0 keV
while little variability is detected below 1 keV. The spectral analysis of the
three phases reveals the presence of a hot component at 3.0 keV that adds up 
to two relatively cold components at 0.9 keV and 2.2 keV. 
We explain the smooth variability with the emergence of an extended active region on 
the surface of the primary star due to its fast rotation (v$sin~i\sim315$ km/s). 
We estimate that the region has diameter in
the range $0.5-0.6$ R$_*$. The hard X-ray emission and its variability 
hint a magnetic origin, 
{ as suggested for few other late-O$-$early-B type stars.}   
We also discuss an alternative explanation based on the emergence from 
occultation of a young (5-10 Myr) low mass companion bright and hot in X-rays.} 
   \keywords{stars: activity -- stars: individual (Rho-Ophiuchi) 
-- stars: early-type -- stars: magnetic field -- stars: starspots}

   \maketitle
%

\section{Introduction}
X-ray emission is a common feature among massive stars of spectral type O through early B-type.
In single O-stars, a soft X-ray emission is thought to be generated by shocks in the stellar
winds accelerated by the strong X-UV stellar flux in a non linear mechanism   
called Line De-shadowing Instability (LDI, \citealp{Owocki1988,Feldmeier1997a,
Feldmeier1997b}). In strongly magnetized massive early-type stars,  
winds can be channeled and collide at high Mach numbers generating hard X-ray emission 
\citep{Babel1997, ud-Doula2002}. In binary systems, large-scale shocks associated with 
wind-wind collisions can manifest themselves in additional X-ray emission 
\citep{Stevens1992}.
Examples of variability from colliding winds  on a time scale of a few years synchronized
with the orbital period of O-type binaries are observed in three systems of the Cyg OB2
complex  \citep{Cazorla2014}. 
 

The origin of magnetic fields in O and B stars is thought to reside either 
in a dynamo mechanism at the interface between convective core and radiative 
layers (but it is still difficult to model the buoyancy of the magnetic field),
or it has a fossil origin being the ambient magnetic field from the parent 
cloud trapped and compressed 
during the formation of the star, with some subsequent dynamo driven amplification
\citep{Walder2012}.
In both cases, the magnetic fields likely have a main dipolar component although 
more complex configurations can be present. In this framework, the origin of X-rays 
from stellar winds is less probable in B type stars because of their weaker  
winds with respect to those of O and WR stars. The rate of detection in X-rays
among B stars falls to about 50\% \citep{Wolk2011,Naze2011,Rauw2014} 
and hard X-ray emission in a few cases are
a signature of the presence of strong magnetic fields. 
Significant examples of X-ray emitters among late O- early B type stars are given by $\tau$~Sco 
\citep{Cohen2003},  $\beta$ Crucis \citep{Cohen2008}, 
$\zeta$~Puppis \citep{Naze2013},  and $\theta^1$~Ori~C \citep{Gagne1997}.
The detection of spots on the surface of late O / early B-type stars have been reported 
in X-ray band  by \citet{Gagne1997} on $\theta^1$~Ori~C (O7 V, rotation period $p=15.4$) 
with {\em ROSAT},  and recently by \citet{Fossati2014} in two early B-type stars of the 
NGC~2264 star forming region. 
{ Another case of modulated variability and hard X-ray spectrum in early B star 
which hints a magnetic origin is $\sigma$~Ori~E \citep{Skinner2008}.} 
%
Here we report on X-rays from \rhoOph, { a binary system at a distance of $\sim111$ pc from the Sun,
composed of a B2IV star and a B2V star \citep{Abt2011,VanLeeuwen2007}.} 
The separation bwtween the two stellar components is about { 2.8$\arcsec$ or 310 AU at
a distance of 111 pc}, and the orbital period is $\sim 2000$ years. 
\rhoOph\ sits about $1\deg$ north of the dense core of active star formation L~1688, 
and gives it the name  ``the \rhoOph\ dark cloud complex''. About 300
members in various stages of formation - from Class 0/I to Class II and III -
belong to L~1688 { \citep{,Wilking08}} at an average distance of $120-130$ pc. 
A ring of dust visible in far to mid infrared and with radius 
$\sim40^\prime$ (1.4 pc) surrounds \rhoOph\ and possibly formed by the winds of the 
central binary system. In this area a number of Pre Main Sequence stars have formed likely 
coeval to \rhoOph.
We observed \rhoOph\ with \xmm\ for a duration of $\sim53$ ks in order to identify and
to study this first episode of star formation likely triggered by 
\rhoOph\ system itself. Indeed, we discovered a group of about 25 young low mass stars 
around \rhoOph\ that are older ($\sim5$ Myr, Pillitteri et al., in preparation) 
than the embedded population in L1688 ($<1$ Myr). 
\begin{figure}
\includegraphics[width=\columnwidth]{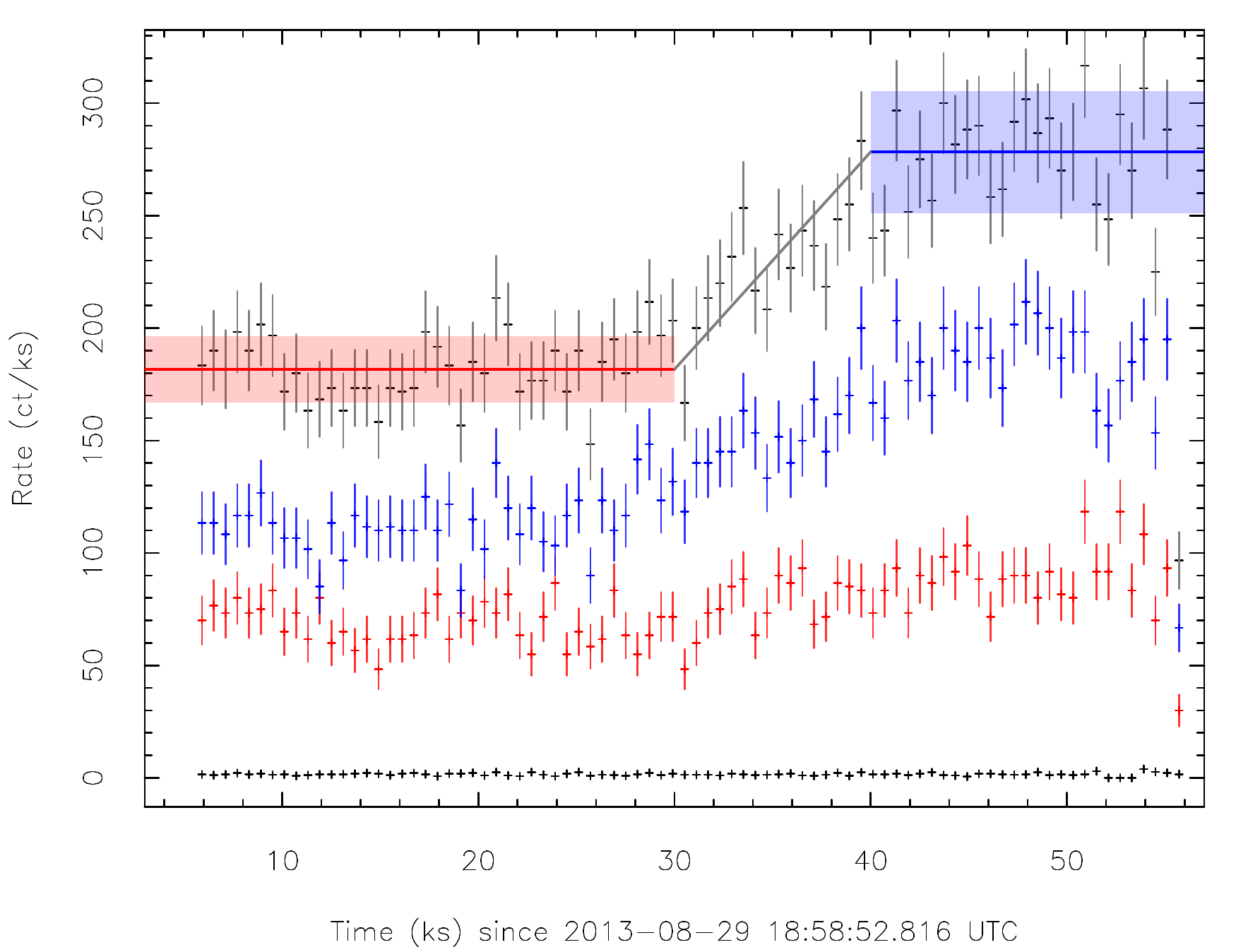}
\caption{\label{lc-rho-oph}
\pn\ light curve of $\rho$ Ophiuchi in broad band (0.3-8 keV), hard band (blue, $>1$ keV), 
soft band (red, $<1.0$ keV), and background. {The events are extracted from a circular region
centered on the X-ray centroid position (see Table \ref{optpos}) and of radius 16$\arcsec$.}  
We observe three phases in \rhoOph's curve:
a low rate phase ($0-30$ ks), a rise phase ($30-40$ ks) and a high rate phase ($>40$ ks).
The rate in the quiescent phase has a median value of $\sim182$ ct/ks with a mean absolute 
deviation (MAD, marked with the red shaded area) of 15 ct/ks. High phase rate has a 
median rate of 280 ct/ks with MAD = 27 ct/ks (blue shaded area). 
The change in the rate is mostly seen above 1 keV, while the soft band rate remains almost 
constant.}
\end{figure}
\section{X-ray light curve and spectra}
{The \xmm\ observation was taken on August 29th 2013, for a nominal duration of $\sim53$ ks. 
The {\em Thick} filter was used for the EPIC exposures in order to prevent damage to
the instruments and to avoid UV contamination from \rhoOph, which has magnitude $V=5.05$.}
 The data sets were analyzed with SAS version 13.5, starting from the Observation Data Files
and reducing them to obtain calibrated event lists in $0.3-8.0$ keV band for MOS and \pn\ cameras, 
including a filtering of events for PATTERN $< 4$ and FLAG $<0$.

\begin{figure}
\includegraphics[width=\columnwidth]{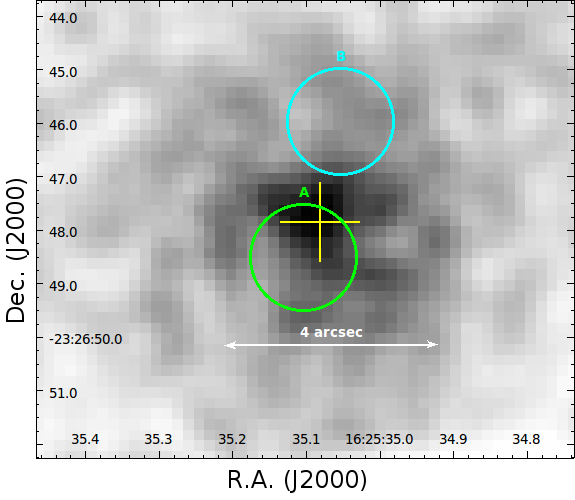}
\caption{\label{zoom-rho-oph}
Mos 1 image and positions of the two components of \rhoOph\ system, A and B.
The original image is binned in blocks of size $0.2\arcsec\times0.2\arcsec$, and smoothed 
with a gaussian kernel of 3 pixels. { The color scale is linear. 
The cross marks the position of the X-ray centroid and gives a visual estimate of 
the positional uncertainty.}
The circles have radii of $1\arcsec$ to account for the precision of the optical positions. 
Although we cannot separate X-ray emission of both components, the centroid of the emission appears
closer to the A component.
 }
\end{figure}

\begin{table}
\caption{\label{optpos} Optical positions of the components of \rhoOph A+B system and of the
X-ray centroid.}
\begin{tabular}{lll}\hline\hline
Object & R.A. (FK5, J2000) & Dec. (J2000) \\\hline
\rhoOph~A & $16^h25^m35.118^s$ & $-23^d26^m49.81^s$ \\
\rhoOph~B & $16^h25^m35.05^s$  & $-23^d26^m46.0^s$ \\
X-ray centroid & $16^h25^m35.2^s$ & $-23^d26^m48.3^s$\\\hline    
\end{tabular}   
\end{table}   
We detected copious X-ray emission from the \rhoOph\ system. 
Given the spatial resolution of the EPIC camera and the errors associated to the positions of 
the X-ray sources we cannot firmly distinguish between the X-ray emission of \rhoOph~A and B (see
Table \ref{optpos} for the positions of the components of \rhoOph\ system).
The {\em Point Spread Function} (PSF) of EPIC is about $4.4\arcsec$ (FWHM), the calibration of absolute
astrometry has nominally zero systematic shift and less than $0.8\arcsec$ of error. 
Relative astrometry of EPIC has uncertainties of less than $1.2\arcsec$. 
The positional uncertainty for \rhoOph\ from the wavelet detection code 
\citep{Damiani97a, Damiani97b} is $<1\arcsec$. We expect that the combination of these
source of astrometric uncertainties amount to $\le2\arcsec$.
A visual inspection of the MOS 1 image shows that the centroid of the
emission is closer to the optical position of \rhoOph~A (Fig. \ref{zoom-rho-oph}), thus
we tentatively assign the X-ray emission and its features to the primary star.

\rhoOph\ has been observed with \xmm\ without interruptions for about 53 ks. 
{The background rate was generally low, but elevated during the last 5 ks. 
Data acquired during the high background interval were filtered out only for 
improving the source detection process and to maximize the signal to noise of faintest
sources. Here we considered the full length exposure.}
In the \pn\ light curve of \rhoOph\ we recognize three phases (Fig. \ref{lc-rho-oph}). 
{ A first phase of relatively low rate (median \pn\ rate: 182 ct/ks, 
median MOS~1 rate: $62\pm12$ ct/ks; median MOS~2 rate: $63\pm12$ ct/ks, 
all rates given in 0.3-8.0 keV band), 
a second phase of gradual increase of the rate lasting about 10 ks, 
then a phase of high rate  (median \pn\ rate 280 ct/ks;
median MOS~1 rate: $95\pm 12$ ct/ks; median MOS~2 rate: $91\pm 6$ ct/ks, all rates given in 
0.3-8.0 keV band).} 
Small flickering at level of one sigma is visible on top of the average flux during 
the whole observation. 
The light curves in soft ($0.3-1.0$ keV) and hard ($>1$ keV) bands show that the increase 
of flux is smooth essentially observed in the hard band.

We have analyzed the MOS~1, MOS~2 and \pn\ spectra from the time interval in $0-30$~ks, $30-40$~ks and 
$40-60$~ks in order to identify spectral changes. The best fit parameters are shown
in Table \ref{bestfit}.
During the initial low rate phase, a best fit to the spectrum is obtained with a sum of 
two thermal APEC components plus global absorption. The temperatures are $kT_1 = 0.9$ keV 
and $kT_2 = 2.2$ keV with the hot component accounting for
about 30\% of the total plasma emission measure. 

For the best fit of the rise and the high rate phases, 
we have added a third component to the 2T model 
used for the low rate spectrum. Our hypothesis is that a hot component appears gradually on the
visible face of the star during the rise, and remains visible during the high rate phase. 
The two components of the low 
rate phase were kept fixed while we allowed to vary only the hot component temperature, 
and its normalization. Plasma as hot as 3.0-3.4 keV is found during the rise
and the high rate phase. The temperatures of the third hot component, estimated during rise and high 
rate intervals, are consistent with each other within one $\sigma$ reinforcing the scenario of 
a hotter region emerging in the visible part of the primary star. 
The change of flux is manifested by a gradual increase of the 
emission measure of the hot plasma component, which increases from $5.1\times10^{52}$ cm$^{-3}$ to 
$9.7\times10^{52}$ cm$^{-3}$. This is fully compatible with the hypothesis of a region brighter 
and hotter than the average stellar surface that gradually appears on view during the rise 
of the count rate. The flux varies by a factor 40\% going from low rate to high rate phases.

Compared to the B stars in Carina region, \rhoOph\ is quite peculiar with regard to its 
X-ray properties. \citet{Gagne2011} find that B stars in Carina with X-ray luminosity higher than 
$L_X \ge 10^{31}$ erg s$^{-1}$ have a hard spectrum ($kT > 1$ keV) while B stars with $L_X < 10^{31}$ 
erg s$^{-1}$ have softer spectra ($kT < 0.6$ keV). Despite its relatively low X-ray luminosity 
($L_X\sim10^{30}$ erg s$^{-1}$), the spectrum of \rhoOph\ is quite hard during the low rate phase 
($kT_2\sim 2$ keV) and becomes even harder during the rise and high rate phase ($kT_3 \sim 3.0-3.4$ keV).   

{ \citet{Stelzer2005} report the case of three late O / early B type stars from the COUP sample that
exhibit periodic X-ray variability phased with the rotational stellar period. The hardness ratios
of these stars are periodic as well with the hardest spectrum corresponding to the highest rate as
we find in \rhoOph.}
 
\begin{figure}
\includegraphics[width=1.05\columnwidth]{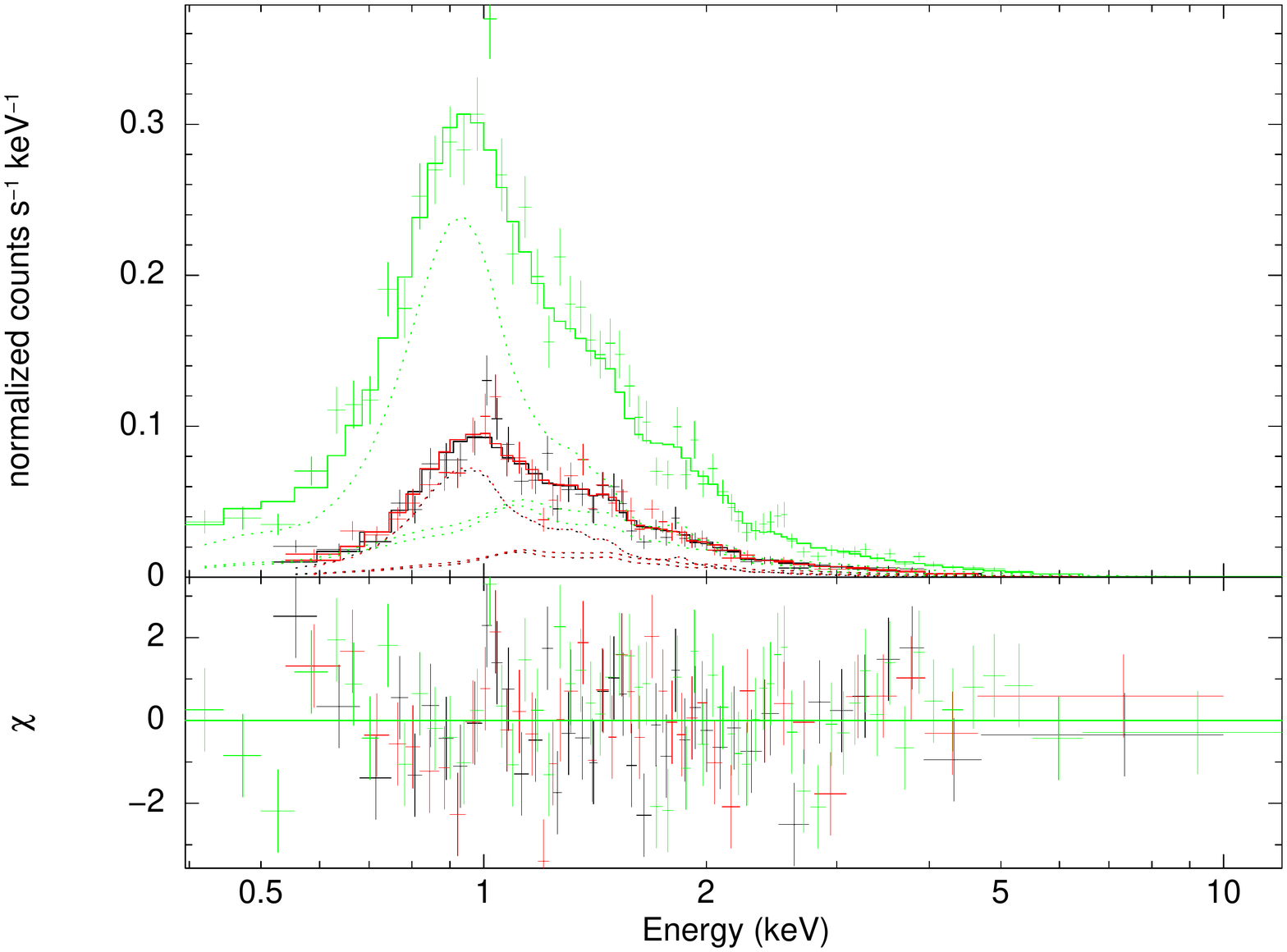}
\includegraphics[width=1.05\columnwidth]{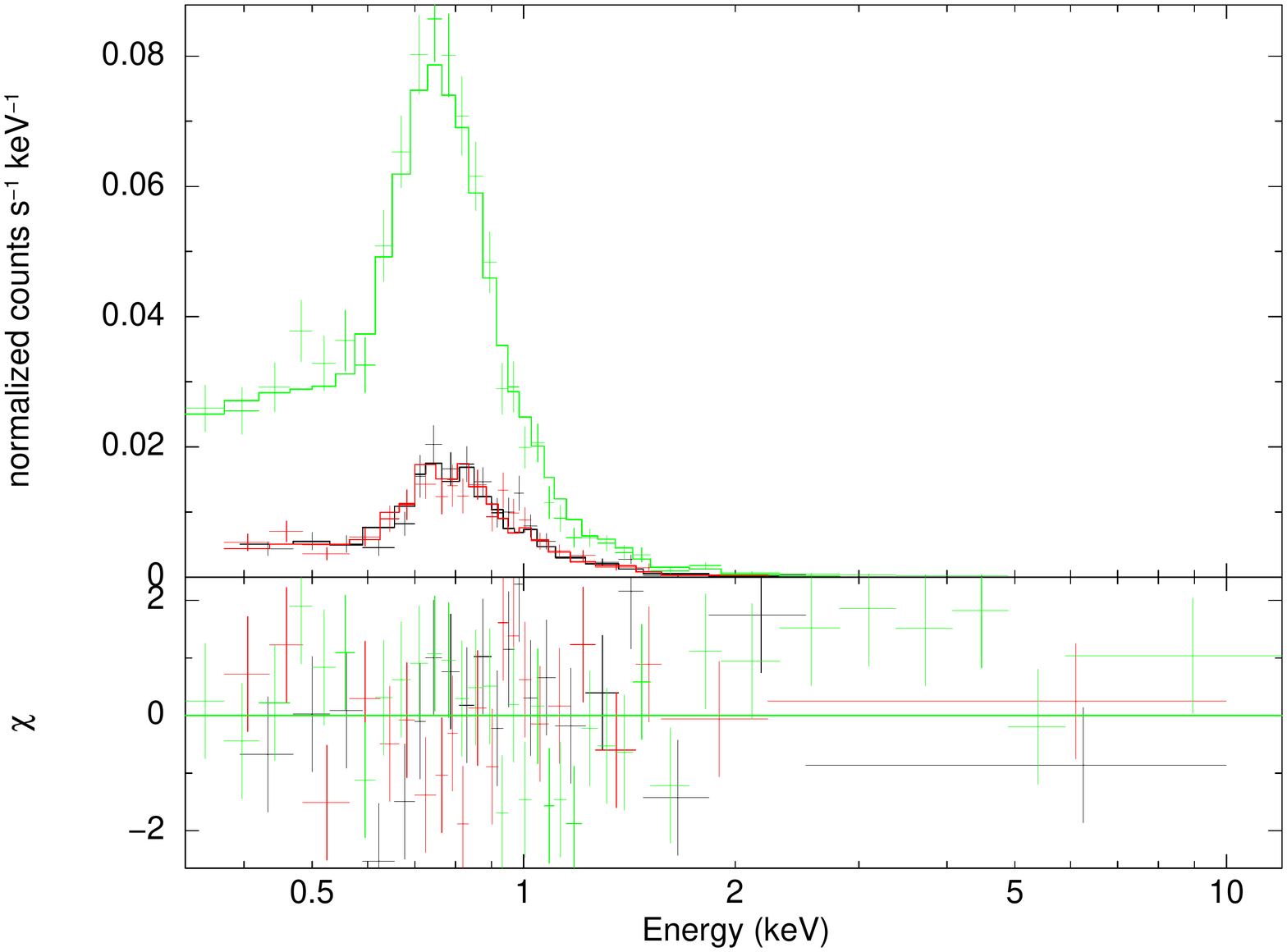}
\caption{\label{spec} Top figure: EPIC MOS~1 (black), MOS~2 (red) and {\em pn} (green) spectra of \rhoOph\ 
during the high rate phase (top panel) and $\chi^2$ terms (bottom panel). 
Dotted lines The APEC components are plotted along with the data. A high temperature of $\sim 3$ keV is 
detected during the rise and the high rate phase.
Bottom figure: same plot for HIP 100751. Notice the harder spectrum of \rhoOph\ compared to HIP 100751.
{ The spectra are accumulated from the same events selected for the light curve 
(see Fig. \ref{lc-rho-oph}).} }
\end{figure}

For further comparison, we have analyzed the EPIC spectra of HIP 100751 ($\alpha$~Pavonis) 
which is also a B2IV type star in a binary system and recently observed by \xmm\ 
(ObsId: 0690680201, PI: W. Waldron).
The light curve of   HIP 100751 shows brief impulsive variability, more typical of young solar mass stars,
and perhaps originated by the unresolved stellar companion.
The spectrum of HIP 100751 is quite soft (Fig. \ref{spec}, bottom panel), 
the best fit is obtained with two thermal APEC components 
at $kT_1=0.4$ keV and $kT_2=0.8$ keV, with the cold component emission measure being the 70\% 
of the total emission measure, and the total X-ray luminosity is $2.6 \times 10^{28}$ erg s$^{-1}$
in 0.3-8.0 keV band. We conclude that \rhoOph\ is thus more luminous than HIP 100751 and it has a harder 
spectrum, reinforcing the peculiarity of \rhoOph.   

\begin{table}[t]
\caption{\label{bestfit} Parameters from model best fit to the \pn\ spectra in the three
different temporal phases. Parameters kept fixed in rise and high rate phases are
enclosed in braces.} 
\centering
\resizebox{0.99\columnwidth}{!}{
\begin{tabular}{l|rrr}
  \hline \hline
 	& \multicolumn{3}{c}{Phase:} \\
Parameters:	& Low Rate	& Rise	& High Rate \\
   \hline
$\chi^2$& 124.1		& 89.4	& 208.6 \\
D.o.F.  	& 134		& 99	& 147   \\
$N_H$ (cm$^{-2}$) 	& 0.33$\pm$0.02 & $0.32\pm0.03$	& $0.31\pm0.07$	 \\
$kT_1$ (keV) 	& 0.9$\pm$0.03	& (0.9)	&  (0.9)	 \\
Abundance ($Z/Z\odot$)	& 0.13$\pm0.02$	& (0.13)	&  (0.13)	 \\
$EM_1$ (10$^{52}$cm$^{-3}$)	& 21.8$\pm$0.7	& (21.8)	&  (21.8)	 \\
$kT_2$ (keV) 		& 2.1$\pm0.4$	& (2.1)	& (2.1)	 \\
$EM_2$ (10$^{52}$cm$^{-3}$)		& 7.8$\pm$2.9	& (7.8)	& (7.8)	 \\
$kT_3$ (keV) 		& --		& 3.4 $\pm0.5$ & 3.0$\pm0.3$	 \\
$EM_3$ (10$^{52}$ cm$^{-3}$)		& --		& $5.1\pm$0.7	& $9.7\pm0.7$ \\
Flux (10$^{-12}$ erg s$^{-1}$ cm$^{-2}$)	& 1.50		& 1.88	&  2.2\\
   \hline
\end{tabular}
}
\end{table}

\section{Discussion and conclusions}
The X-ray variability and the hardness of the X-ray spectrum of \rhoOph\ make this star
interesting in the context of X-ray emission from massive stars. 
We have observed \rhoOph\ in X-rays with \xmm\ and we found a smooth increase of the
X-ray flux above 1 keV. We tentatively assign the whole X-ray emission 
to the primary component given the shorter distance of the X-ray flux centroid from its 
optical position. However, the low spatial resolution of the EPIC camera compared to the 
separation of the two components of the system ($\sim2.8\arcsec$) does not determine 
precisely the source of X-rays. 

We observe a low rate phase followed by a smooth rise and a high rate state.
A first possible interpretation of this behaviour is that the increase of rate as due to a 
hot active region emerging from the limb on the stellar surface because of 
the rotation of the star. We measured the rotational broadening $v$~sin$i$ of \rhoOph\ A
from the width of the He~I line at 6678\AA\ in an archival UVES spectrum, 
obtaining a value of  v$sin~i \sim315$ km/s in agreement with literature values 
($v$~sin$i\sim300$ km/s). 
Given a duration of the rise phase of about 10 ks, we estimated a  
diameter of the active region on the equator of the star of about $3.15\times 10^6$ km.
This size corresponds to about $0.5-0.6 \times R_*$, and thus the active region 
is quite large. 
The active region could be at the equator or at any higher stellar latitude and its origin is
likely due to a strong magnetic field. If the magnetic field 
has a main dipolar shape then an oblique magnetic rotator is involved and the spot is at high
latitude. Were the spot at the equator, a more complex toroidal
field has to be involved. 
\rhoOph\ is not the first case of spotted B stars as we observed before, 
as spots have been identified in other O-late and B-early stars, 
e.g. two young B2 stars in NGC~2264 \citep{Fossati2014}. In these latter cases, the magnetic
origin of the surface spots is inferred by means of  spectropolarimetric observations, 
with magnetic fields of a few hundreds of Gauss 
of intensity and configurations as simple and double dipoles, respectively. 
\citet{Fossati2014} discuss the origin of such magnetic fields, preferring the 
hypothesis of a merging event during the star formation rather than a fossil origin.
The relevance of \rhoOph\ in this context is that its close distance allows
detailed further investigations. 

Another  explanation for the smooth variability is that a low mass unknown companion 
of \rhoOph~A appears to view  and bearing an additional flux with 
intrinsic harder emission. 
The companion could  emerge from behind the primary accounting for the rise phase
during the two limb contacts. Given the difference of flux between
low and high state, its luminosity should be $L_X\sim 1.1\times10^{30}$ erg s$^{-1}$, 
which is on the median X-ray luminosity 
of young stars bright in X-rays that we detected around \rhoOph\  
(median $L_X \sim 1.25\times10^{30}$ erg s$^{-1}$). 
In a simplistic hypothesis, at 1 AU the keplerian orbital velocity 
 is $\sim89$ km s$^{-1}$, and considering
a 10 ks time  for the full disk appearance, a radius of 0.6 R$_\odot$ is derived. 
The radius of the companion is inversely proportional to 
the period (i.e. if $r = 0.5 R\odot$, P $\simeq 50$ days), while its mass does not play
a role. Very little constraint on mass and radius comes from the value of X-ray luminosity 
that this companion should bear ($L_X= 1.1\times10^{30}$ erg s$^{-1}$). 
Because of the saturation of $L_X$ expected at this age 
(5-10 Myr) for $G-K$ type stars \citep{Nicola2003,Jardine99,Favata2003} the presence of 
a relatively small companion like a K early type star is plausible. 
A very close companion coeval of \rhoOph\ should experience its
X-ray luminosity enhanced by the likely spin-orbit locked rotation and thus values of
$L_X \sim 10^{30}$ erg s$^{-1}$ could be observed in such object.
A main drawback for this scenario is the lack of enhancement of the soft emission (e.g. below
1 keV) due to the unknown companion (see Fig. \ref{lc-rho-oph}).
It is quite unlikely that the spectrum of this stellar object is hard or it is heavily
absorbed by intervening material. The absorption toward \rhoOph\ is indeed quite moderate
($N_H \sim 3.4\times10^{21}$ cm$^{-2}$) and soft emission during the low state interval 
is detected. 

{ \rhoOph\ could be similar to the very young system of Oph~S~1, which is likely constituted 
by a B type star and an unknown low mass companion. Oph~S~1 is known to emit variable X-rays
and is a compact radio source \citep{Andre1988,Hamaguchi2003,Gagne2004}. However, 
unlike \rhoOph, in Oph~S~1 the thick cocoon that shrouds the central objects 
could have a significant role in shaping the radio and X-ray emission, likely because of 
the infalling circumstellar material or the interaction of the stellar winds and stellar 
magnetic field with the surrounding material.} 

We find even less likely an X-ray variability arising from colliding winds between the 
two components
of the system, given the weak winds expected from the two B2 stars and their separation.
A marked dropoff of the X-ray flux is observed around spectral type B1 marking a change in the
mechanisms of production of X-rays from stellar winds \citep{Cassinelli1994,Cohen1997}.
\rhoOph\ does not fit into this observational evidence.  
In this respect, given its relatively close distance, \rhoOph\ is one of the best target 
for further investigation of any weak X-ray emission from colliding winds 
originated from the region between the two components of the system.  


In summary, we have presented results of the first pointed X-ray observation of the young 
B-star binary system \rhoOph\ A+B which reveal clear variability on 
a timescale of a few hours 
that is likely associated with magnetic activity, 
with some other low-level flickering possibly present on shorter timescales.
These results, along with detections of similar X-ray variability in a few other young B stars, 
lend support to the presence of magnetic fields in early B stars. However, in the case of 
Rho Oph AB, X-ray observations at higher spatial resolution are needed to more accurately 
determine the position of the X-ray centroid relative to the two closely-spaced B star components 
and to confirm that the X-ray emission is entirely due to the primary.
%

\begin{acknowledgements}
IP acknowledges dr. Mario Guarcello and dr. Javier Lopez-Santiago 
for the helpful discussions on the topics of this paper.
IP acknowledges financial support of the European Union
under the project ``Astronomy Fellowships in Italy" (AstroFit).
S.J.W. was supported by NASA contract NAS8-03060.
\end{acknowledgements}


\end{document}